\documentclass[
   ,final            % use final for the camera ready runs
  ]
  {aipproc}

  \def\be{\begin{equation}}
      
  \def\beq{\begin{eqnarray}}  \def\eeq{\end{eqnarray}}
  \def\be{\begin{equation}}   \def\ee{\end{equation}}

\layoutstyle{6x9}
\usepackage{hyperref}
\begin{document}

\title{Decay properties of charm and beauty open flavour mesons}

\classification{12.39.Pn; 12.40.Yx; 13.20.Fc; 13.20.He} \keywords
{charmed meson , Bottom meson, masses and decay constants}

\author{Ajay Kumar Rai}{
  address={Applied Physics Department, Faculty of Technology
\& Engineering,\\ The M S University of Baroda, Vadodara 390 001,
INDIA.}, }

\author{P C Vinodkumar}{
address={Department of Physics, Sardar Patel University, Vallabh
Vidyanagar 388 120, INDIA.}, }

\begin{abstract}
The masses of $S$ and $P$ states, pseudoscalar and vector decay
constants, leptonic, semileptonic decay widths of charm (D) and
beauty (B) open flavour mesons have been computed in the framework
of Coulomb and power potential of the form
$V(r)=-\frac{\alpha_c}{r}+A r^{\nu}$. The results are compared with
other theoretical as well as experimental results.
\end{abstract}

\maketitle

%%%%%%%%%%%%%%%%%%%%%%%%%%%%%%%%%%%%%%%%%%%%
%% MAINMATTER
%%%%%%%%%%%%%%%%%%%%%%%%%%%%%%%%%%%%%%%%%%%%

\section{INTRODUCTION}
The investigation of the decay properties of charm and beauty open
flavour mesons (D, and B) are interesting and important from the
point of view of recent experimental developments in light-heavy
flavour physics (CELO, Belle, BaBar etc.) \cite{pdg2006}. These
experimental collaborations have contributed many recent charmed and
beauty spectroscopy results \cite{MBiasini2005,Antimo2006}. Future
programs of these collaborations will provide rich spectroscopy and
decay properties of strange and non-strange charmed and beauty
mesons. Theoretically, charm provides a unique window to study QCD
process as its mass lies some where in light quark chiral limit and
the limit where the heavy quark effective theory is applicable. Due
to its intermediate mass scale, charmed mesons are being produced
almost all the high energy experiments. It is also expected that the
charmed mesons productions can be studied through the decay modes of
beauty mesons. High statistics samples of B-decays which allows to
study meson spectroscopy from light flavour to charm and charmonium
mesons are expected from the super B-factories
\cite{MBiasini2005,Antimo2006}. Non- perturbative models have been
successfully employed for the understanding of the decay rates and
other details concerning its radial wave functions and form factor
for the study of leptonic and semileptonic decay of charm and beauty
mesons \cite{W Y Wang2003}.

%In this paper we present our results on the investigations of the
%mass spectrum (S and P-wave), decay constants, leptonic,
%semileptonic, non-leptonic decays and life time of these mesons
%based on a general potential model [5].

\section{Theoretical framework}
For the light-heavy flavour bound system we treat the heavy-quark
(Q=c, b) non-relativistically and the light-quark (q = u, d, s)
relativistically. The Hamiltonian for the case be written as
\cite{akr2002, akr2005}\be \label{eq:hamiltonian}
H=M+\frac{p^2}{2m}+\sqrt{p^2+m^2}+V(r)+V_{S_{\bar {Q}} \cdot
S_q}(r) + V_{{L} \cdot S}(r)\ee

Where M is the heavy quark mass, m is the light quark mass, p is the
relativistic momentum of each quark, V(r) is the quark- antiquark
potential $V_{S_{\bar {Q}} \cdot S_q}(r)$ and  $V_{{L} \cdot S}(r)$
are the spin-spin and spin orbital interactions. Here we consider
 \be V(r)=\frac{-\alpha_c}{r}+A r^{\nu}\ee
where $\alpha_c=\frac {4}{3}\alpha_s$, $\alpha_s$ being the strong
running coupling constant, A and $\nu$ are the  potential
parameter. The value of A is different for different $\nu$. The
spin dependent part of the usual one gluon exchange potential
(OGEP) between the quark antiquark for computing the hyperfine and
spin-orbit shifting of S and P states are given by
 \cite{SSGershtein1995}

\be V_{S_{\bar {Q}} \cdot S_q}(r)=\frac{2}{3}
\frac{\alpha_c}{M_{\bar{Q}} m_q} \ \vec{S_{\bar {Q}}} \cdot
\vec{S_q} \ 4\pi \delta(\vec{r}), \ \ \  V_{{L} \cdot S}(r)=
\frac{\alpha_c}{M_{\bar{Q}} m_q}\ \frac{\vec{L} \cdot
\vec{S}}{r^3} \ee We employ the hydrogenic trial wave function and
use the virial theorem, \be\label{eq:virial}
 \left<\frac{p^2}{2M}\right>=\frac{1}{2}{\left<\frac{rdV}{dr}\right>}\ee
to get the energy expression for the ground state as
 \be E(\mu, \nu)=M+\frac{\mu^2}{8M}+\frac{\mu^2}{8m}-
 \frac{1.75}{8^3} \frac{\mu^5}{m^3}+\frac{1}{2} \left(\frac{A\ \Gamma(\nu+3)}{\mu^{\nu}}-\alpha_c \mu\right)\ee
where $\mu$ is the wave function parameter determined using the
viral theorem of Eq(\ref{eq:virial}). For the excited states
appropriate hydrogenic wave function are being used in the present
study. The parameters used here are $m_{u/d}=0.360\ GeV$,
$m_{c}=1.41 \ GeV$, $m_{b}=4.88 \ GeV$, $\alpha_c=0.48$ (for D
meson) and $\alpha_c=0.36$ (for B \ meson). The mass spectrum of D
and B mesons are tabulated in Tables (\ref{tab:1}) and
(\ref{tab:2}) alongwith the experimental and other theoretical
results.

\begin{table}
\caption{S-Wave and P-Wave Masses (in $GeV$) of $D$ meson}
\label{tab:1}
\begin{tabular}{ccccccccccc}
%\multicolumn{10}{}{}\\
\hline $\nu$&$1^1S_0$&$1\ ^3S_1$&1\ $^3 P_0$&1\ $^3
P_1$&$1^1P_1$&1\ $^3 P_2$&$2^1S_0$&2\ $^3S_1$&$3^1S_0$&3\ $^3S_1$ \\
\hline
0.5&1.889&2.005&2.203&2.217&2.232&2.246&2.267&2.291&2.478&2.494\\
0.7&1.855&2.019&2.257&2.281&2.305&2.329&2.342&2.399&2.643&2.675\\
0.9&1.823&2.038&2.305&2.341&2.376&2.411&2.399&2.489&2.784&2.839\\
1.0&1.806&2.046&2.324&2.365&2.407&2.449&2.416&2.523&2.832&2.903\\
1.1&1.790&2.057&2.342&2.390&2.440&2.489&2.423&2.554&2.851&2.950\\
1.3&1.755&2.082&2.371&2.437&2.503&2.569&2.481&2.577&-&-\\
1.5&1.715&2.133&-&-&-&&-&-&-&-\\
\hline
Expt.\cite{pdg2006}&1.864&2.006&&&&&&2.54\cite{glwang2006}&&\\
\hline
Ebert\cite{ebert1998}&1.875&2.009&2.414&2.438&2.459&2.501&2.579&2.629&&\\
\hline
Pandya\cite{jnp2001}&1.815&1.909&2.385&2.417&2.449&2.481&2.653&2.690&3.162&3.175\\
\hline
\end{tabular}\\
\end{table}

\begin{table}
\caption{S-Wave and P-Wave Masses (in $GeV$) of $B$ meson}
\label{tab:2}
\begin{tabular}{ccccccccccc}
%\multicolumn{10}{}{}\\
\hline $\nu$&$1^1S_0$&$1\ ^3S_1$&1\ $^3 P_0$&1\ $^3
P_1$&$1^1P_1$&1\ $^3 P_2$&$2^1S_0$&2\ $^3S_1$&$3^1S_0$&3\ $^3S_1$ \\
\hline
0.5&5.299&5.320&5.500&5.503&5.505&5.508&5.547&5.552&5.698&5.701\\
0.7&5.293&5.324&5.553&5.557&5.561&5.566&5.615&5.625&5.830&5.856\\
0.9&5.285&5.326&5.599&5.605&5.611&5.617&5.668&5.684&5.942&5.952\\
1.0&5.281&5.327&5.620&5.627&6.635&6.642&5.689&5.709&5.987&6.000\\
1.1&5.277&5.329&5.640&5.649&5.658&5.666&5.705&5.729&6.019&6.035\\
1.3&5.267&5.333&5.679&5.690&5.702&5.713&5.711&5.746&-&-\\
1.5&5.254&5.337&-&-&-&&-&-&-&-\\
\hline
Expt.\cite{pdg2006}&5.279&5.324&&&&&&5.842\cite{glwang2006}&&\\
\hline
Ebert\cite{ebert1998}&5.285&5.324&5.719&5.738&5.733&5.757&5.883&5.898&-&-\\
\hline
\end{tabular}\\
\end{table}

\section{Pseudoscalar and vector mesons decay constants and weak decays}
The decay constants of the meson are very important parameters in
the determination of the leptonic, non-leptonic weak decay processes
and CKM matrix. It is related to the wave function at the origin
through Van-Royen-Weisskoff formula \cite{vrw1967}. Incorporating a
first order QCD correction factor, we compute them using the
relation \cite{BF1995}
 \be f^2_{P/V}=\frac{12\left|\Psi_{P/V}(0)\right|^2}{M_{P/V}}
C^2(\alpha_s), \ \ \ \ \textmd{Where} \ \
C^2(\alpha_s)=1-\frac{\alpha_s}{\pi}\left[2-\frac{M_Q-m_q}{M_Q+m_q}\
ln\frac{M_Q}{m_Q} \right]\ee where $M_{P/V}$ is the ground state
mass of the pseudoscalar/vector meson. \\
For weak decays of B and D mesons, we employ the spectator model
\cite{alv1999,akr2006}. In the spectator approximation the
inclusive widths of b and c quarks  are given as
 \be
 \Gamma(b \rightarrow X)= \frac{ 9 \ G^2_F \left|V_{Q\bar{Q}}\right|^2
 m^5_b}{192 \pi^3}, \ \ \
 \Gamma(c \rightarrow X)= \frac{ 5 \ G^2_F \left|V_{Q\bar{q}}\right|^2
 m^5_c}{192 \pi^3}  \ee
and width of the annihilation channel is computed using the
expression given by \cite{alv1999} \be\label{ttlannrest}
  \Gamma(Anni)= \frac{G^2_F}{8 \pi}\left|V_{Q\bar{q}}\right|^2 f^2_{P} M_{P}
\sum_i m^2_i\left(1- \frac{m^2_i}{M^2_{P}}\right)^2 \ . \ C_i, \ee
where $C_i=1$ for the $\tau \nu_{\tau}$ channel and
$C_i=3\left|V_{Q\bar{q}}\right|^2$ for $Q\bar{q}$, and $m_i$ is
the mass of the heaviest fermions. Here$\left|V_{Q\bar{q}}\right|$
and $\left|V_{Q\bar{Q}}\right|$ are CKM Matrix, taken from
\cite{pdg2006}. The total width of the corresponding decay is the
addition of partial widths $\Gamma(total)$ $=$ $\Gamma(Q
\rightarrow X)+ \Gamma(Anni)$. The total widths and lifetime are
not changing much with variation of $\nu$. The pseudoscalar,
vector decay constants, total width and lifetime of D and B mesons
are listed
in Tables (\ref{tab:3})and (\ref{tab:4}).\\

{\bf Acknowledgement:} One of the author P. C. Vinodkumar
acknowledge the financial support from Department of Science and
Technology, Government of India under a Major research project
SR/S2/HEP-20/2006.

\begin{table}
\caption{Decay constants ($f_P$ $\&$ $f_V$) and lifetime of D
meson (in MeV). } \label{tab:3}
\begin{tabular}{ccccccccc}
\hline Models&$R_p(0)$&$R_v(0)$&$f_P$
&$f_P(cor.)$&$f_{V}$&$f_{V}(cor.)$&$\Gamma(total)$&$\tau$ \\
&$GeV^{3/2}$&$GeV^{3/2}$&$MeV$&$MeV$&$MeV$&$MeV$&$10^{-4} eV$&$ps$\\
\hline
$\nu$\ =    \ 0.5&0.4216&0.4493&300&259&310&268&6.088&1.081\\
\ \ \ \ \ \ \ 0.7&0.4973&0.5460&357&308&376&324&&\\
\ \ \ \ \ \ \ 0.9&0.5633&0.6397&408&352&438&378&&\\
\ \ \ \ \ \ \ 1.0&0.5926&0.6854&431&372&468&404&&\\
\ \ \ \ \ \ \ 1.1&0.6214&0.7337&454&392&500&432&&\\
\ \ \ \ \ \ \ 1.3&0.6758&0.8379&498&430&568&490&&\\
\ \ \ \ \ \ \ 1.5&0.7428&1.0038&552&477&672&580&6.331&1.040\\
\hline
Others&&&&230\cite{gcvetic2004}&&339$\pm$22\cite{glwang2006}&&1.040$\pm$\\
&&&&195$\pm$20\cite{aapenin2002}&&&&0.007\cite{pdg2006}\\
&&&&243$\pm$25\cite{ebert2002}&&&&\\
\hline
\end{tabular}
\end{table}

\begin{table}
\caption{Decay constants ($f_P$ $\&$ $f_V$)and lifetime of B meson
(in MeV).} \label{tab:4}
\begin{tabular}{ccccccccc}
\hline Models&$R_p(0)$&$R_v(0)$&$f_P$
&$f_P(cor.)$&$f_{V}$&$f_{V}(cor.)$&$\Gamma(total)$&$\tau$ \\
&$GeV^{3/2}$&$GeV^{3/2}$&$MeV$&$MeV$&$MeV$&$MeV$&$10^{-4} eV$&$ps$\\
\hline
$\nu$\ =\ 0.5&0.3918&0.3934&166&170&167&170&10.005&0.655\\
\ \ \ \ \ \ \ 0.7&0.4758&0.4751&201&205&202&206&&\\
\ \ \ \ \ \ \ 0.9&0.5365&0.5508&232&237&233&238&&\\
\ \ \ \ \ \ \ 1.0&0.5822&0.5874&248&253&249&254&&\\
\ \ \ \ \ \ \ 1.1&0.6176&0.6238&263&268&264&270&&\\
\ \ \ \ \ \ \ 1.3&0.6907&0.6994&294&300&296&302&&\\
\ \ \ \ \ \ \ 1.5&0.7750&7874&330&337&333&340&10.060&0.654\\
\hline
Others&&&&196\cite{gcvetic2004}&&238$\pm$18\cite{glwang2006}&&1.638$\pm$\\
&&&&206$\pm$20\cite{aapenin2002}&&&&0.011\cite{pdg2006}\\
&&&&178$\pm$15\cite{ebert2002}&&&&\\
\hline
\end{tabular}
\end{table}

\end{document}